\documentclass{article}

\usepackage{spconf,amsmath,graphicx,bm,booktabs,verbatim,amsfonts,makecell}

\title{UNSUPERVISED PRE-TRAINING FOR SEQUENCE TO SEQUENCE SPEECH RECOGNITION}
\name{Zhiyun Fan$^{1,2}$, Shiyu Zhou$^1$, Bo Xu$^1$}
\address{
	$^1$Institute of Automation, Chinese Academy of Sciences, China\\
	$^2$University of Chinese Academy of Sciences, China\\
	\{fanzhiyun2017, zhoushiyu2013, xubo\}@ia.ac.cn}

\begin{document}
\ninept
\maketitle
%
\begin{abstract}
This paper proposes a novel approach to pre-train encoder-decoder sequence-to-sequence (seq2seq) model with unpaired speech and transcripts respectively. Our pre-training method is divided into two stages, named acoustic pre-trianing and linguistic pre-training. In the acoustic pre-training stage, we use a large amount of speech to pre-train the encoder by predicting masked speech feature chunks with its context. In the linguistic pre-training stage, we generate synthesized speech from a large number of transcripts using a single-speaker text to speech (TTS) system, and use the synthesized paired data to pre-train decoder. This two-stage pre-training method integrates rich acoustic and linguistic knowledge into seq2seq model, which will benefit downstream automatic speech recognition (ASR) tasks. The unsupervised pre-training is finished on AISHELL-2 dataset and we apply the pre-trained model to multiple paired data ratios of AISHELL-1 and HKUST.
 We obtain relative character error rate reduction (CERR) from 38.24\% to 7.88\% on AISHELL-1 and from 12.00\% to 1.20\% on HKUST.  Besides, we apply our pre-trained model to a cross-lingual case with CALLHOME dataset. For all six languages in CALLHOME dataset, our pre-training method makes model outperform baseline  consistently. 
\end{abstract}
\section{Introduction}
\label{sec:intro}
There have been growing interests in building an end-to-end (E2E) speech recognition system, which directly transduces an input sequence of acoustic features to an output sequence of tokens. Because of a single all-neural structure, E2E system has several advantages, including simpler training process  and joint optimization among components. Prominent E2E models include: (a) connectionist temporal classification (CTC)  \cite{graves2006connectionist, graves2014towards}, (b) attention based encoder-decoder networks \cite{cho2014learning, bahdanau2014neural, bahdanau2016end, chorowski2015attention, vaswani2017attention}, and (c) recurrent neural network transducer (RNN-T) \cite{graves2012sequence}.  

Although E2E models are powerful, they still meet the issue that training process is very hungry for human-transcripted supervised data. Unfortunately, the supervised data collection is a time-consuming and expensive process, especially for speech-to-text data. Comparing with supervised data, unpaired data (speech and text) is much easier to collect. 
So a lot of unsupervised and semi-supervised methods \cite{chen2019completely, chung2018unsupervised, karita2018semi, li2019semi} are proposed to easy the dependence. 

Recently, the release of BERT (Bidirectional Encoder Representation from Transformer) provides us a new way to utilize unpaired data by pre-training. BERT is a bidirectional variant of Transformer networks trained to jointly predict a masked word from its context and to classify whether two sentences are consecutive or not. Then the pre-trained model can swiftly adapt for downstream tasks by fine-tuning just one additional output layer.
According to \cite{jawahar2019does}, BERT can capture the structural information about language contained in text-only data by pre-training and these semantic information is helpful to downstream tasks. Intuitively, similar to text, speech as another carrier of semantic information can be processed in a similar way.  

In this paper, we propose a two-stage unsupervised pre-training for attention-based encoder-decoder framework.  Two pre-training stages are used to extract acoustic and linguistic information with speech and transcripts respectively. In the first stage, we pre-train the encoder with a large amount of unlabeled speech data. We mask some continuous feature chunks in each sequence at random, and use context of these masked feature chunks to predict them. With this acoustic pre-training, encoder obtains a good initial parameters. However the parameter of attention and decoder are still scratch. This is why BERT is not well suited for tasks that require decoding text, e.g, automatic speech recognition (ASR). To remedy this deficiency, we turn to a second pre-training stage.  In the  second stage, we generate speech for a large number of transcripts with a trained text to speech (TTS) \cite{zou2019boosting} system and use these synthesized data to optimize the attention and decoder part. Although the acoustic information of synthesized data is monotonous, there is rich linguistic information contained in these transcripts, which is useful for downstream ASR task.  

All of our experiments are conducted on Transformer \cite{vaswani2017attention} belonging to encoder-decoder framework. After two-stage pre-training with AISHELL-2 (remove test set of AISHELL-1), we apply the pre-trained model to several downstream tasks. We obtain relative character error rate reduction (CERR) from $38.24\%$ to $7.88\%$ on the test set of AISHELL-1. 
For HKUST dataset, we obtain relative CERR from $12.00\%$ to $1.20\%$. In addition, we apply our pre-trained model to CALLHOME dataset, which contains six low-source languages. Experiments demostrate obviously promoting effect on all six languages, which indicates that our pre-training method works well even if in a cross-lingual case.
\section{RELATED WORK}
The most related work to this paper is BERT \cite{devlin2018bert}, which is a bidirectional language representation model. BERT is trained to capture useful representations by predicting masked tokens with their context and classifying the relationship between two sentences. Thus when fine-tuning on downstream tasks, model can converge faster and better by using the learned representation than initializing with scratch. As for ASR task, we propose a pre-training approach to capture useful representaion contained in speech. Our pre-training policy is different from BERT mainly in two aspects: 1. We mask a continuous feature vector sequence rather than discrete tokens. Unlike text which can be broken into character or word units relatively easily, speech features are continuous, and the neighboring frames are similar. So only predicting discrete frames is too easy for neural networks. 2. We don't use the next sentence prediction mentioned in \cite{devlin2018bert}. Because the seq2seq system decodes one sentence at a time, there is no needs to understand sentence relationships. 

Another well-known pre-training technique is restricted Boltzmann machine (RBM), which is a particular type of Markov Random Field (MRF) that has one layer of binary stochastic hidden units and one layer of binary stochastic visible units \cite{mohamed2009deep}. Using RBM, deep neural networks (DNN) can be pre-trained layer by layer in an unsupervised way. Comparing our unsupervised pre-training with RBM-based pre-training, there are three differences: 1. Our method uses a joint optimization instead of a layer-by-layer way, which means our pre-training process is simpler. 2. This paper utilizes both speech and transcripts to conduct pre-training, while \cite{mohamed2009deep} only uses speech. 3. The framework is different. \cite{mohamed2009deep} only pre-trains acoustic model in hybrid systems while this work pre-trains the whole E2E model.
\section{UNSUPERVISED PRE-TRAINING}
\label{sec:format}
\begin{figure}[htb]
		\centering
		\centerline{\includegraphics[width=8.0cm]{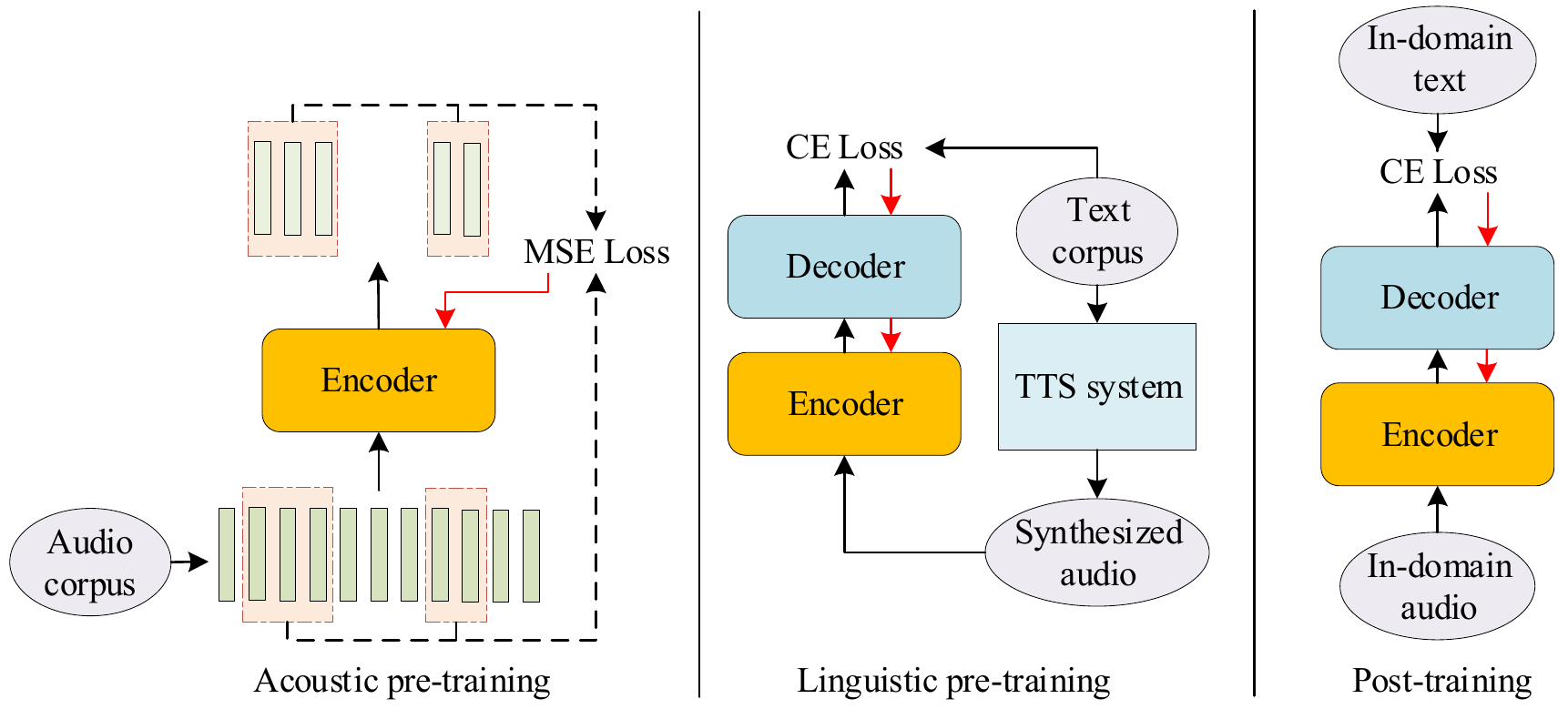}}
		\caption{A schematic representation of our unsupervised pre-training method}
	\label{ph1}
\end{figure}

In this paper, we use Transformer as the study platform to investigate our unsupervised pre-training method, which consists of two stages named acoustic pre-training and linguistic pre-training. This section discusses the details of the two-stage pre-training method. As a attention-based encoder-decoder model, Transformer can be divided into three parts, that is, encoder, decoder and attention. The acoustic pre-training aims to integrate useful representation contained in speech into encoder by predicting some masked feature chunks in speech feature sequence. In the linguistic pre-training, we use a trained TTS system to generate speech from a large number of transcripts. Using these synthesized paired data, decoder can obtain rich linguistic information. Fig.\,\ref{ph1} illustrates the details of our unsupervised pre-training method. 
\subsection{Acoustic pre-training}
\label{acoustic}
The left side of Fig.\,\ref{ph1} illustrates the acoustic pre-training. To pre-train the encoder, we mask some continuous feature sequences of input $x=(x_1,x_2,...,x_T)$ along the time steps. The size and position of these masked feature chunks are sampled randomly. We set the masked value in chunks to zero, and only predict these masked features rather than reconstruct the entire input. The details of our strategy which is used to create masked feature sequence for each sentence are shown as follows:
\begin{itemize}
\item $K$ masked chunks: We mask $K$ feature chunks, which are represented with dashed box in the left side of Fig.\,\ref{ph1}. Firstly, $K$ time points are chosen along the time steps $(0, T)$ as the centers of these feature chunks for each sentence. $T$ is the length of the whole feature sequence. And chunk size $w$ is sampled from a uniform distribution from 0 to $W$. Here $W$ represents the max length that can be masked during training. 
\begin{equation}
  c_i \sim uniform(0, T), i \in (1, 2, ..., K)
  \label{eq1}
\end{equation}
\begin{equation}
  w \sim uniform(0, W)
  \label{eq2}
\end{equation}
We denote the indexes of feature vectors in the i-th masked feature chunk as $f_i$:
\begin{equation}
s_i = max(0, c_i-w), e_i = min(c_i+w, T)
\label{eq3}
\end{equation}
\begin{equation}
  f_i = [s_i, e_i], i \in (1, 2, ..., K)
  \label{eq4}
\end{equation}
\item 80\% of the time: For each feaure chunk, there is an 80\% chance the values in chunk are masked to zero.
\begin{equation}
 \setlength{\abovedisplayskip}{4pt}
 \setlength{\belowdisplayskip}{4pt}
  x_t = \bm{0}, t \in f_i
  \label{eq5}
\end{equation}
\item  20\% of the time: The feature chunks are not always masekd. There is a 20\% chance the values in chunk stay the same. The purpose of this is to bias the representation towards the actual speech feature sequence. 
\end{itemize}

The encoder reads a masked sequence of d-dimensional feature vector $x'=(x'_1, x'_2, ..., x'_T)$, and transforms it to a higher-order representation $h=(h_1,h_2,...,h_T)$. At the top of encoder, a extra linear layer projects hidden feature to the same dimension as input feature. This final output sequence can be regarded as a hypothesis of input feature sequence, denoted as $\hat{x}=(\hat{x}_1,\hat{x}_2,...,\hat{x}_T)$.

Instead of reconstructing the entire input, we only predict these masked features. We use mean square error (MSE) loss to conduct training. 
\begin{equation}
 \setlength{\abovedisplayskip}{4pt}
 \setlength{\belowdisplayskip}{4pt}
  Loss=\frac{1}{BK}\sum_{b=1}^{B}\sum_{i=1}^{K}\sum_{t\in f_i}\left\|x_{b,t}-\hat{x}_{b,t}\right\|^2
  \label{eq5}
\end{equation}
Where the subscript $(b)$ indicates the b-th example in a batch which contains B examples. The $f_i$ is the indexes of feature vectors in the i-th masked feature chunk.  
\subsection{Linguistic pre-training}
\label{ssec:acoustic}
ASR is a speech-to-text task, and seq2seq model consists of encoder and decoder. Our acoustic pre-training which is used to pre-train encoder is not enough for a encoder-deocder framework like Transformer. This section proposes a linguistic pre-training to remedy this defect with a large number of transcripts. 

A common approach to leverage text-only data is training a language model (LM), and fusing the output of decoder and the pre-trained LM. Deep fusion \cite{gulcehre2015using} and shallow fusion \cite{chorowski2016towards} are two approaches to integrate LM into E2E system. Although these fusion approaches have shown improvement to E2E system, the drawback of them is the extra LM increases complexity to the system. In this paper, we use a trained TTS system to generate speech for a large number of transcripts, which transfers text-only data to paired data. Then we use these synthesized paired data to train a Transformer whose encoder is initialized with acoustic pre-training mentioned in the section \ref{acoustic}. This supervised training uses cross entropy (CE) loss, and backpropagation updates the whole model, including encoder, decoder and attention. The whole process is illustrated in the middle of Fig.\,\ref{ph1}. This pre-traing stage aims to integrate linguistic information into seq2seq system. Although the acoustic information of these synthesized paired data is monotonous, the linguistic information contained in these transcripts is rich. 
%
%
\begin{table*}[!htbp ]
	\caption{CER[\%]/WER[\%] performance of our two-stage pre-training method on six languages of CALLHOME.}
	\label{tab:3}
	\vspace{2mm}
	\centering
	\begin{tabular}{p{0.9cm}<{\centering}p{1.3cm}<{\centering}p{1.1cm}<{\centering}p{1.1cm}<{\centering}p{1.1cm}<{\centering}p{1.1cm}<{\centering}p{1.1cm}<{\centering}p{1.1cm}<{\centering}}
		\toprule
		Exp &Models & EN & MA & JA& SP& GE & AR\\
		\hline
		C0 & scratch & -  &  - &  -  &   -   &  -  & -\\
		C1 & C0 + M1 &50.55  & 49.33 & 46.41&  62.08  &  57.03 & 56.62 \\ 
		\bottomrule
	\end{tabular}
\end{table*}


Comparing our approach with extra LM \cite{gulcehre2015using, chorowski2016towards} and BERT \cite{devlin2018bert}, using synthesized paired data to integrate linguistic information mainly has two advantages: 1. Decoder of transformer stacks identity blocks which contain three sublayers, that is, self-attention, encoder-decoder attention and feed forward network. BERT-initialized decoder can only initialize two sublayers, and the encoder-decoder attention is still initialized randomly. However the encoder-decoder attention represents the alignment between speech and text. Supervised training with synthesized paired data can help system to capture the linguistic information and alignment between speech and text simultaneously. 2. The supervised training with synthesized paired data integrates linguistic information into decoder of Transformer without extra LM, which leads to a more simple model structure and reduces the complexity of system.

\subsection{In-domain Post-training}
Using our unsupervised pre-training, seq2seq model extracts rich acoustic and linguistic representation that are useful to downstream ASR tasks. After pre-training, seq2seq model still needs fine-tuning with in-domain paired data. On the right side of Fig.\,\ref{ph1}, we illustrate this post-training process. In this stage, supervised training is conducted by CE loss and model is initialized with the average of last checkpoints in linguistic pre-training. The softmax layer is reinitialized randomly, and the number of output units depends on training set. During fine-tuning, we update the parameters of encoder, decoder and attention simultaneously.
\section{EXPERIMENTS}
\label{sec:pagestyle}
\subsection{Data Sets}
We evaluate our approach on four public ASR datasets, that is, AISHELL-2 \cite{du2018aishell}, AISHELL-1 \cite{bu2017aishell}, HKUST \cite{liu2006hkust} and CALLHOME \cite{zhou2017multilingual}. AISHELL-2, a Mandarin ASR dataset, which contains about 1000 hours speech-to-text data, is used to conduct our unsupervised pre-training.  It should be noted that AISHELL-1 is a subset of AISHELL-2. We remove these sentences appearring in the test set of AISHELL-1 from AISHELL-2. The rest paired data of AISHELL-2 is splitted into speech and transcripts, which are used to conduct our acoustic and linguistic pre-training respectively.
AISHELL-1, a 178 hours speech dataset, performs as paired in-domain data for supervised post-training. Because of inclusive relationship between AISHELL-1 and AISHELL-2, AISHELL-1 is used to simulate the case that the distribution of pre-training data and in-domain data is almost consistent. HKUST is a spontaneous speech corpus (201 hours), whose distribution is quite different from AISHELL-2. CALLHOME is a multi-language dataset, which contains six languages including Mandarin (MA), English (EN), Japanese (JA), Arabic (AR), German (GE) and Spanish (SP). The details of this dataset can be found in \cite{zhou2017multilingual}. We use CALLHOME dataset to explore the effect of our approach on a cross-langual case. 

The synthesized data used in the linguistic pre-training stage is generated from $99392$ transcripts (remove $7176$ transcripts in the test set of AISHELL-1) in AISHELL-2, and the duration of synthesized audio is up to $800$ hours.  The structure and training details of the speech synthesis system can be found in \cite{zou2019boosting}.  
\subsection{Modeling and training}
All the acoustic features used in this paper are 80-dimensional log-Mel filter-bank features, computed with a 25 ms window and shifted every 10 ms. The raw features are normalized via mean subtraction and variance normalization per speaker side. Before sending into model, the features are firstly stacked with 3 frames to the left and then down-sampled to 33.3 Hz frame rate. We mask stacked $320$ fbank feature sequence with $K=2$ and $W=10$.
 
Transformer used in this paper contains 6 encoder and 6 decoder blocks, with a per-block configuration of $d_{model}=512$, $16$ attention heads, and $2048$ feed forward inner-layer dimension. 
In the linguistic pre-training, we use $3961$ characters appearing in the AISHELL-2 and $4$ extra tokens, including an unknown token ($<$UNK$>$), a padding token ($<$PAD$>$), and sentence start and end tokens ($<$S$>$/$<$$\backslash$S$>$) as output units. In the post-training, the softmax layer is reinitialized randomly. We use $4230$ and $3896$ characters plus $4$ extra takens as output units for AISHELL-1 and HKUST respectively. For CALLHOME dataset, we use word pieces plus 4 extra tokens as output units. And the EN, MA, GE, AR, JA, SP have $552$, $3956$, $567$, $555$, $3090$, $568$ output units respectively. For more details of Transformer, we refer readers to \cite{vaswani2017attention}.

 During both pre-training and post-training we use the Adam optimizer with ${\beta_1}=0.9 $, $\beta_2=0.98$, $\varepsilon={10^{ - 9}}$ and alter the learning rate over the course of training. In the linguistic pre-training and post-training, the label smoothing of value $\varepsilon_{ls}  = 0.1$ is employed \cite{szegedy2016rethinking}. And the last $20$ checkpoints are averaged for inference. For evaluation, we use beam search with a beam size of $13$ and length penalty $\alpha=0.6$. We use CER to evaluate all Chinese and Japanese datasets in this paper, and word erro rate (WER) for other languages.

Our acoustic pre-training only uses speech of AISHELL-2 to train the encoder of Transformer, which is stacked an extra full connected (FC) layer that projects dimension of features to $320$. After the acoustic pre-training stage, we discard the extra FC layer and only keep the parameters of encoder, denoted as M0. In linguistic pre-training, we use the transcripts of AISHELL-2 to train a whole Transformer whose encoder is initialized with M0. We denote this model as M1. To conduct ablation study in the section \ref{ablation}, we apply our linguistic pre-training to a Transformer initialized randomly and denote this model as M2. In addition, we pre-train a Transformer initialized randomly with real paired AISHELL-2, denoted as M3. In the post-training, the softmax layer of models that are initialized with M1, M2 and M3 is initialized randomly.
\section{RESULTS}
\label{sec:typestyle}
\subsection{Comparison with baseline models}
\begin{figure}[htb]
	\hfill
	\begin{minipage}[b]{0.48\linewidth}
		\centering
		\centerline{\includegraphics[width=3.0cm]{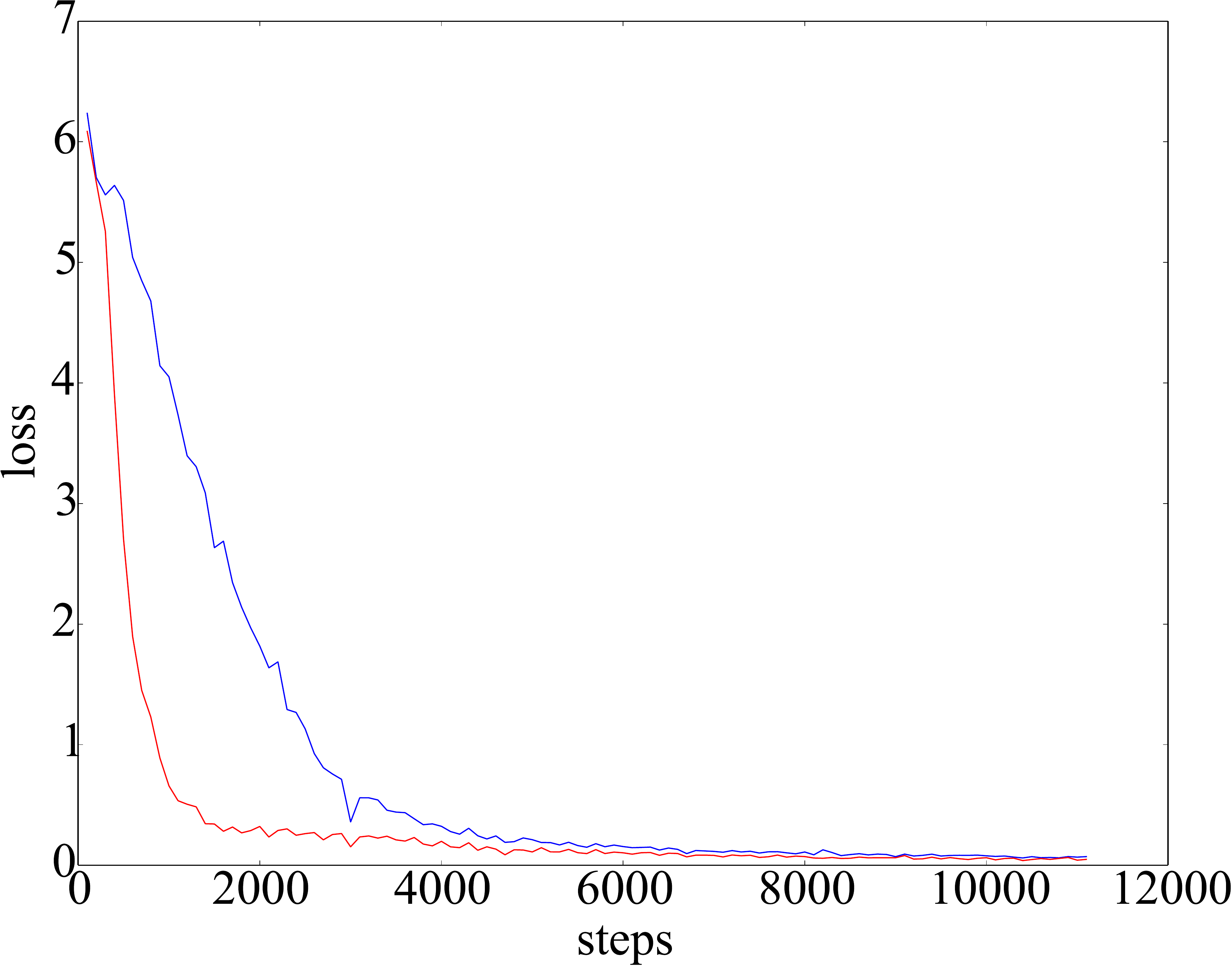}}
		\centerline{(a) 20 hours}\medskip
	\end{minipage}
	\hfill
	\begin{minipage}[b]{0.48\linewidth}
		\centering
		\centerline{\includegraphics[width=3.0cm]{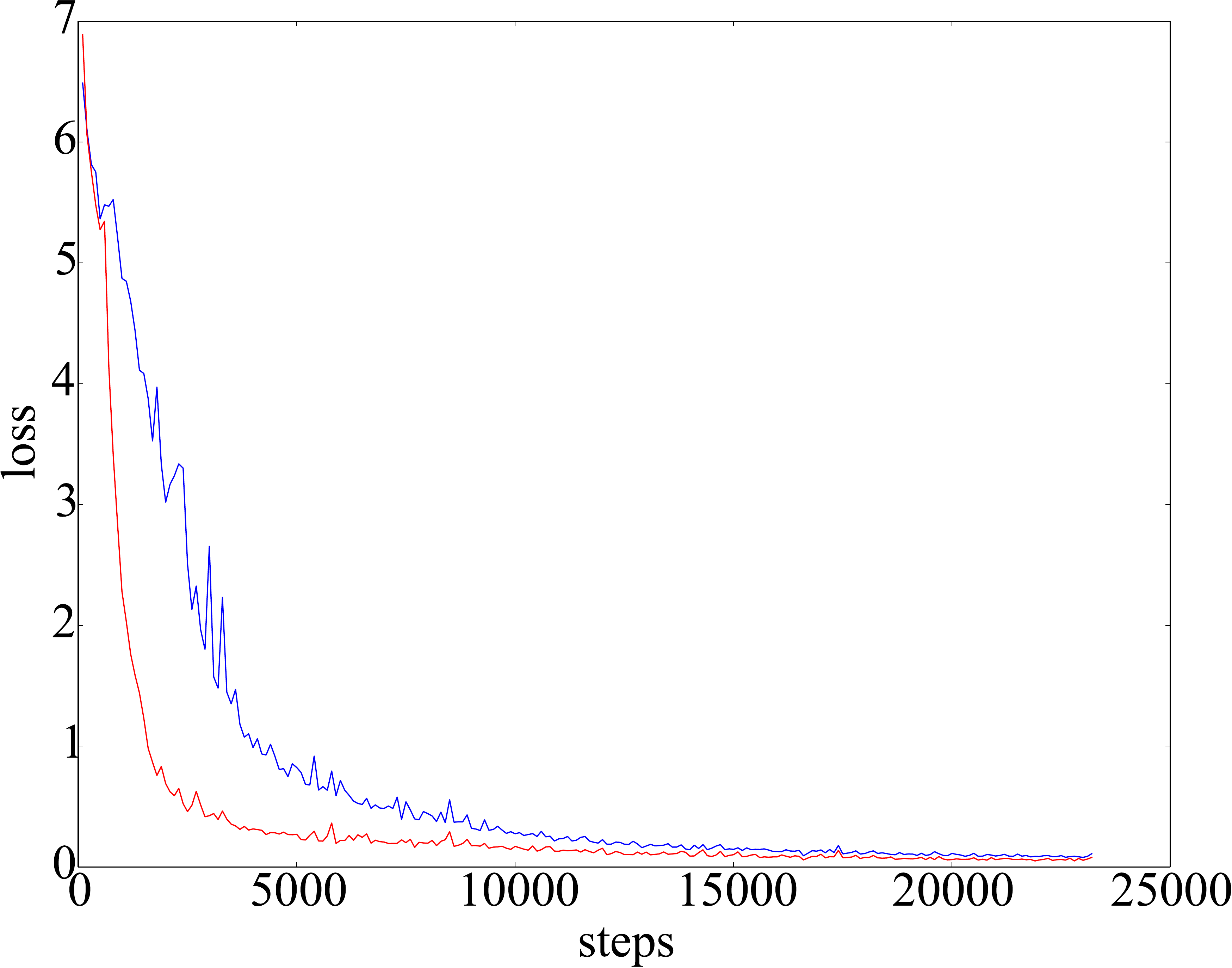}}
		\centerline{(b) 178 hours}\medskip
	\end{minipage}
	\caption{Loss curve of in-domain post-training with 20 hours and 178 hours of AISHELL-1 .}
	\label{conver}

\end{figure}

In this section, we evaluate our approach on AISHELL-1, whose distribution is consistent with AISHELL-2. We use 20 hours, 40 hours, 89 hours and 178 hours of AISHELL-1 as in-domain paired data for post-training respectively. Table \ref{tab:1} summarizes the CER on the test set of AISHELL-1. A0 is the baseline model initialized with scratch. Initialized with our unsupervised pre-trained model (M2), A1 obtains a relative CERR  from 38.24\% to 7.88\% than baseline system (A0) on the four ratios of AISHELL-1, which indicates that our unsupervised pre-training benefits downstream ASR tasks. And it is obvious that the smaller the amount of in-domain data is, the more significant the improvement is. Besides, models initialized with our unsupervised pre-training method converge consistently faster than randomly initialized baseline. Fig.\,\ref{conver} illustrates the loss curve of $20$ hours and $178$ hours cases. In another two case, the loss curves are similar. Expertiments A4 are initialized with a supervised pre-training model M3. If baseline system (A0) is a lower bounds of our unsupervised pre-training, A4 can be regarded as upper bounds. From A1 and A4, our unsupervised pre-training has  been pretty  close to the supervised pre-training.
\begin{table}[th]
	\caption{CER[\%] performance of our unsupervised pre-training method on AISHELL-1.}
	\label{tab:1}
	\vspace{2mm}
	\centering
	\begin{tabular}{cccccc}
		\toprule
		Exp &Models & 10h & 20h & 89h & 178h \\
		\hline
		A0 & scratch & 32.77 & 21.22& 11.04&7.87 \\
		A1 & A0 + M1 &\textbf{20.24}	 &\textbf{15.03}	&\textbf{8.98}&	\textbf{7.25}  \\ 
		A2 & A0 + M0  & 25.20 &	16.59	&9.19&	7.45 \\
		A3 & A0 + M2 &23.31 &	16.70&	9.43&	7.56\\
		\hline
		A4 & A0 + M3 & 18.08	& 13.12& 8.16 &6.70 \\
		\bottomrule
	\end{tabular}
\end{table}


\subsection{Ablation study}
\label{ablation}
In this section, we use several ablation studies to demonstrate the effect of acoustic and linguistic pre-training respectively. A2 and A3 in the Table \ref{tab:1} eliminate the effects of linguistic pre-training and acoustic pre-training respectively. A2 obtains relative CERR from 23.10\% to 5.34\% with only acoustic pre-training and A3  obtains relavtive CERR from 28.87\% to 3.94\% with only linguistic pre-training. It means that two stages of unsupervised pre-training can boost the downstream ASR task independently. And when the amount of available paired data for ASR task is very small (10 hours), the linguistic pre-training is more useful than acoustic pre-training. With the amount of paired data increasing, the acoustic pre-training starts to play a more important role.  
\subsection{Mismatch case}
The distribution of AISHELL-1 and AISHELL-2 is similar. But in most case, the distribution of paired data used in post-training is very different from the data used in pre-training procedure. In this section, we evaluate our approach on one fourth, half, three fourths and all of HKUST dataset respectively, which is very different from AISHELL-2 used in pre-training. Comparing B1 with B0 in Table \ref{tab:2}, we obtain relative CERR from 12.00\% to 1.20\%, which indicates that our acoustic pre-training boosts downstream ASR task even if there is mismatch between unpaired pre-training data and in-domain data.
The results of B3 is much worse than B1, and it indicates that in such a case (there is distribution mismatch between pre-training data and in-domain data), acoustic pre-training are more useful than linguistic pre-training. Comparing B2 with B1, we find that linguistic pre-training may bring negative effects when we have applyed acoustic pre-training in such a mismatch case.
\begin{table}[h]
	\caption{CER[\%] performance of our unsupervised pre-training method on HKUST.}
	\label{tab:2}
	\vspace{2mm}
	\centering
	\begin{tabular}{cccccc}
		\toprule
		Exp &Models & 1/4& 1/2& 3/4& ALL \\
		\hline
		B0 & scratch & 43.09 &33.46   &	30.18	&26.56\\
		B1 & B0 + M0  &  \textbf{37.92}   &  \textbf{31.80} & 29.49 &    \textbf{26.24}  \\
		B2 & B0 + M1 &38.05 &	32.07&	\textbf{29.28} &	26.32 \\ 
		B3 & B0 + M2 & 40.29  &     33.16&    30.10 & 26.51 \\
		\bottomrule
	\end{tabular}
\end{table}
\subsection{Cross-lingual case}

For some languages, even unlabeled data is hard to collect. In this case, we may not have huge amount of speech to carry on our unsupervised pre-training. Inspired by \cite{schultz2001language, yu2009cross}, we consider to use unpaired data from other language. In this section, we apply model pre-trained with AISHELL-2 (M1) to six ASR tasks for six languages in CALLHOME dataset. For all six languages, our baseline systems without any pre-trained model (C0) don't converge. Initialized with M1, experiments C1 converge better cross all six languages. It indicates that our unsupervised pre-training approach works well even if the downstream ASR task is for a language which is different from the one used in pre-training. We owe this improvement in the cross-lingual case to the sharing occurring at a certain level of acoustic units such as mono-phones or states \cite{schultz2001language, yu2009cross}.



\section{Conclusion}
\label{sec:copyright}
In this paper, we investigate the use of unpaired speech and transcripts to conduct a two-stage unsupervised pre-training. Experiments demonstrate that these models, which leverage large-scale two-stage pre-training, outperform those that only used in-domain data. And we verify the effectiveness of our approach when there is mismatch between pre-training and post-training data. Furthermore, we show that our pre-training method works well in the cross-lingual case. In the future, we plan to apply our approach to larger datasets and investigate more efficient unsupervied pre-training method.
\section{Acknowledgements}
The authors would like to thank Jingqing Ruan and Yuxiang Zou for helpful discussions.

\vfill\pagebreak

\bibliographystyle{IEEEbib}
\bibliography{strings,refs}

\begin{thebibliography}{10}

\bibitem{graves2006connectionist}
Alex Graves, Santiago Fern{\'a}ndez, Faustino Gomez, and J{\"u}rgen
  Schmidhuber,
\newblock ``Connectionist temporal classification: labelling unsegmented
  sequence data with recurrent neural networks,''
\newblock in {\em Proceedings of the 23rd international conference on Machine
  learning}. ACM, 2006, pp. 369--376.

\bibitem{graves2014towards}
Alex Graves and Navdeep Jaitly,
\newblock ``Towards end-to-end speech recognition with recurrent neural
  networks,''
\newblock in {\em International conference on machine learning}, 2014, pp.
  1764--1772.

\bibitem{cho2014learning}
Kyunghyun Cho, Bart Van~Merri{\"e}nboer, Caglar Gulcehre, Dzmitry Bahdanau,
  Fethi Bougares, Holger Schwenk, and Yoshua Bengio,
\newblock ``Learning phrase representations using rnn encoder-decoder for
  statistical machine translation,''
\newblock {\em arXiv preprint arXiv:1406.1078}, 2014.

\bibitem{bahdanau2014neural}
Dzmitry Bahdanau, Kyunghyun Cho, and Yoshua Bengio,
\newblock ``Neural machine translation by jointly learning to align and
  translate,''
\newblock {\em arXiv preprint arXiv:1409.0473}, 2014.

\bibitem{bahdanau2016end}
Dzmitry Bahdanau, Jan Chorowski, Dmitriy Serdyuk, Philemon Brakel, and Yoshua
  Bengio,
\newblock ``End-to-end attention-based large vocabulary speech recognition,''
\newblock in {\em 2016 IEEE international conference on acoustics, speech and
  signal processing (ICASSP)}. IEEE, 2016, pp. 4945--4949.

\bibitem{chorowski2015attention}
Jan~K Chorowski, Dzmitry Bahdanau, Dmitriy Serdyuk, Kyunghyun Cho, and Yoshua
  Bengio,
\newblock ``Attention-based models for speech recognition,''
\newblock in {\em Advances in neural information processing systems}, 2015, pp.
  577--585.

\bibitem{vaswani2017attention}
Ashish Vaswani, Noam Shazeer, Niki Parmar, Jakob Uszkoreit, Llion Jones,
  Aidan~N Gomez, {\L}ukasz Kaiser, and Illia Polosukhin,
\newblock ``Attention is all you need,''
\newblock in {\em Advances in neural information processing systems}, 2017, pp.
  5998--6008.

\bibitem{graves2012sequence}
Alex Graves,
\newblock ``Sequence transduction with recurrent neural networks,''
\newblock {\em arXiv preprint arXiv:1211.3711}, 2012.

\bibitem{chen2019completely}
Kuan-Yu Chen, Che-Ping Tsai, Da-Rong Liu, Hung-Yi Lee, and Lin-shan Lee,
\newblock ``Completely unsupervised phoneme recognition by a generative
  adversarial network harmonized with iteratively refined hidden markov
  models,''
\newblock {\em arXiv preprint arXiv:1904.04100}, 2019.

\bibitem{chung2018unsupervised}
Yu-An Chung, Wei-Hung Weng, Schrasing Tong, and James Glass,
\newblock ``Unsupervised cross-modal alignment of speech and text embedding
  spaces,''
\newblock in {\em Advances in Neural Information Processing Systems}, 2018, pp.
  7354--7364.

\bibitem{karita2018semi}
Shigeki Karita, Shinji Watanabe, Tomoharu Iwata, Atsunori Ogawa, and Marc
  Delcroix,
\newblock ``Semi-supervised end-to-end speech recognition.,''
\newblock in {\em Interspeech}, 2018, pp. 2--6.

\bibitem{li2019semi}
Bo~Li, Tara~N Sainath, Ruoming Pang, and Zelin Wu,
\newblock ``Semi-supervised training for end-to-end models via weak
  distillation,''
\newblock in {\em ICASSP 2019-2019 IEEE International Conference on Acoustics,
  Speech and Signal Processing (ICASSP)}. IEEE, 2019, pp. 2837--2841.

\bibitem{jawahar2019does}
Ganesh Jawahar, Beno{\^\i}t Sagot, Djam{\'e} Seddah, Samuel Unicomb, Gerardo
  I{\~n}iguez, M{\'a}rton Karsai, Yannick L{\'e}o, M{\'a}rton Karsai, Carlos
  Sarraute, {\'E}ric Fleury, et~al.,
\newblock ``What does bert learn about the structure of language?,''
\newblock in {\em 57th Annual Meeting of the Association for Computational
  Linguistics (ACL), Florence, Italy}, 2019.

\bibitem{zou2019boosting}
Yuxiang Zou, Linhao Dong, and Bo~Xu,
\newblock ``Boosting character-based chinese speech synthesis via multi-task
  learning and dictionary tutoring,''
\newblock {\em Proc. Interspeech 2019}, pp. 2055--2059, 2019.

\bibitem{devlin2018bert}
Jacob Devlin, Ming-Wei Chang, Kenton Lee, and Kristina Toutanova,
\newblock ``Bert: Pre-training of deep bidirectional transformers for language
  understanding,''
\newblock {\em arXiv preprint arXiv:1810.04805}, 2018.

\bibitem{mohamed2009deep}
Abdel-rahman Mohamed, George Dahl, and Geoffrey Hinton,
\newblock ``Deep belief networks for phone recognition,''
\newblock in {\em Nips workshop on deep learning for speech recognition and
  related applications}. Vancouver, Canada, 2009, vol.~1, p.~39.

\bibitem{gulcehre2015using}
Caglar Gulcehre, Orhan Firat, Kelvin Xu, Kyunghyun Cho, Loic Barrault, Huei-Chi
  Lin, Fethi Bougares, Holger Schwenk, and Yoshua Bengio,
\newblock ``On using monolingual corpora in neural machine translation,''
\newblock {\em arXiv preprint arXiv:1503.03535}, 2015.

\bibitem{chorowski2016towards}
Jan Chorowski and Navdeep Jaitly,
\newblock ``Towards better decoding and language model integration in sequence
  to sequence models,''
\newblock {\em arXiv preprint arXiv:1612.02695}, 2016.

\bibitem{du2018aishell}
Jiayu Du, Xingyu Na, Xuechen Liu, and Hui Bu,
\newblock ``Aishell-2: Transforming mandarin asr research into industrial
  scale,''
\newblock {\em arXiv preprint arXiv:1808.10583}, 2018.

\bibitem{bu2017aishell}
Hui Bu, Jiayu Du, Xingyu Na, Bengu Wu, and Hao Zheng,
\newblock ``Aishell-1: An open-source mandarin speech corpus and a speech
  recognition baseline,''
\newblock in {\em 2017 20th Conference of the Oriental Chapter of the
  International Coordinating Committee on Speech Databases and Speech I/O
  Systems and Assessment (O-COCOSDA)}. IEEE, 2017, pp. 1--5.

\bibitem{liu2006hkust}
Yi~Liu, Pascale Fung, Yongsheng Yang, Christopher Cieri, Shudong Huang, and
  David Graff,
\newblock ``Hkust/mts: A very large scale mandarin telephone speech corpus,''
\newblock in {\em International Symposium on Chinese Spoken Language
  Processing}. Springer, 2006, pp. 724--735.

\bibitem{zhou2017multilingual}
Shiyu Zhou, Yuanyuan Zhao, Shuang Xu, Bo~Xu, et~al.,
\newblock ``Multilingual recurrent neural networks with residual learning for
  low-resource speech recognition,''
\newblock 2017.

\bibitem{szegedy2016rethinking}
Christian Szegedy, Vincent Vanhoucke, Sergey Ioffe, Jon Shlens, and Zbigniew
  Wojna,
\newblock ``Rethinking the inception architecture for computer vision,''
\newblock in {\em Proceedings of the IEEE conference on computer vision and
  pattern recognition}, 2016, pp. 2818--2826.

\bibitem{schultz2001language}
Tanja Schultz and Alex Waibel,
\newblock ``Language-independent and language-adaptive acoustic modeling for
  speech recognition,''
\newblock {\em Speech Communication}, vol. 35, no. 1-2, pp. 31--51, 2001.

\bibitem{yu2009cross}
Dong Yu, Li~Deng, Peng Liu, Jian Wu, Yifan Gong, and Alex Acero,
\newblock ``Cross-lingual speech recognition under runtime resource
  constraints,''
\newblock in {\em 2009 IEEE International Conference on Acoustics, Speech and
  Signal Processing}. IEEE, 2009, pp. 4193--4196.

\end{thebibliography}

\end{document}